\documentclass[%
letterpaper,
reprint,
superscriptaddress,
frontmatterverbose, 
preprintnumbers,
amsmath,amssymb,amsfonts,
aps, pra,
showkeys,
]{revtex4-2}

\usepackage[per-mode = symbol]{siunitx}
\let\svqty\qty
\usepackage{physics}
\let\qty\svqty
\usepackage{mathtools,mathrsfs,xfrac}
\usepackage{graphicx}% Include figure files
\usepackage[caption=false,position=top]{subfig}
\usepackage{ragged2e}
\usepackage{bm}% bold math
\usepackage{hyperref}% add hypertext capabilities
\allowdisplaybreaks

\makeatother

    \makeatletter
\def\@fnsymbol#1{\ensuremath{\ifcase#1\or \dagger\or *\or \ddagger\or
   \mathsection\or \mathparagraph\or \|\or **\or \dagger\dagger
   \or \ddagger\ddagger \else\@ctrerr\fi}}
    \makeatother

\newcommand{\SGone}{\text{\tiny{SG1}}}
\newcommand{\SGtwo}{\text{\tiny{SG2}}}
\newcommand{\NP}{\text{\tiny{NP}}}

\newcommand{\HFS}{\text{\tiny{HFS}}}
\newcommand{\rabi}{\text{\tiny{Rabi}}}
\newcommand{\subs}{\text{\tiny{\textit{S}}}}
\newcommand{\subi}{\scriptscriptstyle{I}}
\newcommand{\mue}{\vb{\bm{\mu}}_\text{e}}
\newcommand{\opmue}{\hat{\vb{\bm{\mu}}}_\text{e}}

\newcommand{\opmun}{\hat{\vb{\bm{\mu}}}_\text{n}}
\newcommand{\vux}{\vb{e}_{x}}
\newcommand{\vuy}{\vb{e}_{y}}
\newcommand{\vuz}{\vb{e}_{z}}

\begin{document}

\title{Quantum mechanical modeling of the multi-stage Stern--Gerlach experiment conducted by Frisch and Segr\`e}

\author{S.~Süleyman Kahraman}
\thanks{These authors contributed equally.}
\affiliation{Caltech Optical Imaging Laboratory, Andrew and Peggy Cherng Department of Medical Engineering, Department of Electrical Engineering, California Institute of Technology, 1200 E. California Blvd., MC 138-78, Pasadena, CA 91125, USA}

\author{Kelvin Titimbo}
\thanks{These authors contributed equally.}
\affiliation{Caltech Optical Imaging Laboratory, Andrew and Peggy Cherng Department of Medical Engineering, Department of Electrical Engineering, California Institute of Technology, 1200 E. California Blvd., MC 138-78, Pasadena, CA 91125, USA}

\author{Zhe He}
\affiliation{Caltech Optical Imaging Laboratory, Andrew and Peggy Cherng Department of Medical Engineering, Department of Electrical Engineering, California Institute of Technology, 1200 E. California Blvd., MC 138-78, Pasadena, CA 91125, USA}

\author{Jung-Tsung Shen}
\affiliation{Department of Electrical and Systems Engineering, Washington University in St.~Louis, St.~Louis, MO 63130, USA}

\author{Lihong V.~Wang}
\email[Corresponding email:]{lvw@caltech.edu}
\affiliation{Caltech Optical Imaging Laboratory, Andrew and Peggy Cherng Department of Medical Engineering, Department of Electrical Engineering, California Institute of Technology, 1200 E. California Blvd., MC 138-78, Pasadena, CA 91125, USA}

\date{\today}%

\begin{abstract}
The multi-stage Stern--Gerlach experiment conducted by Frisch and Segr\`e includes two cascaded quantum measurements with a nonadiabatic flipper in between. 
The Frisch and Segr\`e experiment has been modeled analytically by Majorana without the nuclear effect and subsequently revised by Rabi with the hyperfine interaction.
However, the theoretical predictions do not match the experimental observation accurately. 
Here, we numerically solve the standard quantum mechanical model, via the von Neumann equation, including the hyperfine interaction for the time evolution of the spin.
Thus far, the coefficients of determination from the standard quantum mechanical model without using free parameters are still low, indicating a mismatch between the theory and the experiment. 
Non-standard variants that improve the match are explored for discussion.
\end{abstract}

\maketitle

\section{Introduction}\label{physys}

The quantum measurement problem tackles the conundrum of wave function collapse and the Stern--Gerlach (SG) experiment is considered as the first observation of a quantum measurement \cite{gers1922,gers1924, sakurai2011,feynman2011,messiah2020}. 
While the SG observation was interpreted as proof of quantization of spin \cite{uhlg1925, schsl2016,castelvecchi2022stern}, cascaded quantum measurements provide a more stringent test of theories \cite{sakurai2011, hei1927}.
Frisch and Segr\`e (FS) conducted the first successful multi-stage SG experiment \cite{fris1933,fris1933ital,gers1922,fris2021} after improving the apparatus from Phipps and Stern \cite{phis1932}.
Even though more recent multi-stage SG experiments have been conducted, they differ in the mechanisms of polarizing, flipping, and analyzing the spin \cite{Ramsey1956, schb1983, sch1982, higr1978, hig1975, mardz2021, macjf2013, Rubin2004}. 
Most experiments designed for precise atomic measurements use a narrow-band resonant (adiabatic) flipper \cite{Ramsey1956} while the FS experiment uses a wide-band nonadiabatic flipper. 

The FS experiment was suggested by Einstein \cite{schsl2016,schtt2019,fris2021}, studied analytically by Majorana \cite{maj1932,Bassani2006} and later by Rabi \cite{rab1936}.
Majorana investigated the nonadiabatic transition of the electron spin through a closed-form analytical solution, which is now widely used to analyze any two-level systems \cite{ivasn2023}. 
Rabi revised Majorana's derivation by adding the hyperfine interaction but still could not predict the experimental observation accurately.  
Despite additional theoretical studies into similar problems involving multilevel nonadiabatic transitions \cite{motr1936, sch1937, ash2016, ivasn2023, ostn1997, carh1985, higrt1977}, an exact solution with the hyperfine interaction included has not been obtained.

Among the more recent multi-stage SG experiments \cite{hig1975, higr1978, sch1982, schb1983, macjf2013, mardz2021}, the study most similar to the FS experiment uses a sequence of coils to obtain the desired magnetic field \cite{sch1982, schb1983}. 
The models in these works not only simplified the mathematical description of the magnetic fields generated by the coils but also fit free parameters to predict the experimental observations. 
We choose to model the FS experiment over other similar experiments because of the simplicity of the nonadiabatic spin flipper and its historical significance.

Here, we numerically simulate the FS experiment using a standard quantum mechanical model via the von Neumann equation without tuning any parameters and compare the outcome with the predictions by both Majorana and Rabi. 
Even though our approach is a standard method of studying such spin systems, our results do not match the experimental observations. 
This discrepancy indicates that either our understanding of the FS experiment is lacking or the standard theoretical model is insufficient. 
Recent studies have modeled the FS experiment using an alternative model called co-quantum dynamics \cite{wan2022, hetk2022, titgk2022} without resorting to free parameters. 
We believe it is essential to bring the FS experiment to the attention of the research community. 

This paper is organized as follows. 
In Sec.~\ref{FSexp}, we present the experimental configuration used by Frisch and Segr\`e to measure the fraction of electron spin flip.
In Sec.~\ref{FStheo}, we introduce the von Neumann equation and the Hamiltonian for the nuclear-electron spin system.
Numerical results for the time evolution of the spins and the final electron spin-flip probability are shown here.
In Sec.~\ref{discussion}, we compare the numerical results with previous solutions.
Finally, Sec.~\ref{concl} is left for conclusions.
Non-standard variants of the quantum mechanical model are explored in the appendices to stimulate discussion.

\section{Description of the Frisch--Segr\`e experiment}\label{FSexp}

\begin{figure}
    \centering
    \includegraphics[width=0.95\linewidth]{ 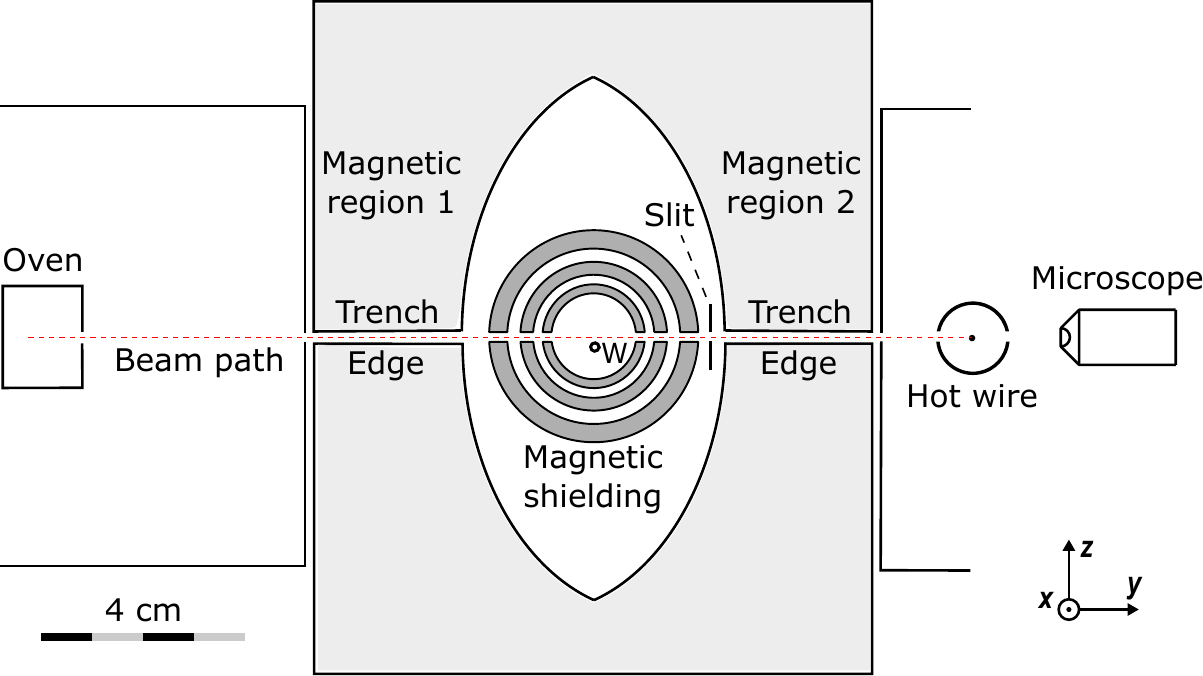}
    \caption{Redrawn schematic of the original setup \cite{fris1933, fris1933ital}. Heated atoms in the oven effuse from a slit. First, the atoms enter magnetic region 1, which acts as SG1. Then, the atoms enter the region with magnetic shielding (i.e., the inner rotation chamber) containing a current-carrying wire W. Next, a slit selects one branch. Magnetic region 2 acts as SG2. The hot wire is scanned vertically to map the strength of the atomic beam along the $z$ axis. The microscope reads the position of the hot wire.}
    \label{fig:FS}
\end{figure}

The schematic of the setup used in the Frisch--Segr\`e experiment \cite{fris1933, fris1933ital} is redrawn in Figure \ref{fig:FS}.
There, magnetic regions 1 and 2 act as Stern--Gerlach apparatuses, SG1 and SG2, respectively, with strong magnetic fields along the $+z$ direction. 
In SG1, stable neutral potassium atoms ($^{39}\mathrm{K}$) effused from the oven are spatially separated by the magnetic field gradient according to the orientation of their electron magnetic moment $\mue$.
The magnetically shielded space containing a current-carrying wire forms the inner rotation (IR) chamber. 
The shielding reduces the fringe fields from the SG magnets down to the remnant field $B_\text{r}=\SI{42}{\micro\tesla}$ aligned with $+\hat{z}$.
Inside the IR chamber, the current-carrying wire placed at a vertical distance $z_\text{a}=\SI{105}{\micro\meter}$ below the atomic beam path creates a cylindrically symmetric magnetic field.
The total magnetic field in the IR chamber equals the superposition of the remnant field and the magnetic field created by the electric current $I_{\text{w}}$ flowing through the wire.
The precise magnetic field outside the IR chamber was not reported \cite{fris1933,fris1933ital}. 
After SG1, the atoms enter the IR chamber; we approximate the motion to be rectilinear and constant along the $y$ axis.
The rectilinear approximation of atomic motion within the IR chamber is acceptable since the total displacement due to the field gradients is negligible, approximately $\SI{1}{\micro\meter}$.
Along the beam path, the magnetic field is given by
\begin{equation}\label{eq:Bexact}
    \vb{B}_{\text{exact}} = \frac{\mu_0 I_\text{w}z_{\text{a}}}{2\pi (y^2+ z_\text{a}^2)}  \,\vuy + \left( B_\text{r} - \frac{\mu_0 I_\text{w} y}{2\pi (y^2+ z_\text{a}^2)}  \right)  \vuz \ ,
\end{equation}
where $\mu_{0}$ is the vacuum permeability; the trajectory of the atom is expressed as $y=vt$, where $v$ is the speed of the atom and the time is set to $t=0$ at the point on the beam path closest to the wire.
The right-handed and unitary vectors $\left\{ \vux, \vuy,\vuz \right\}$ describe the directions of the Cartesian system.

The magnetic field inside the IR chamber has a current-dependent null point below the beam path at coordinates $(0,y_{\NP}, -z_{\text{a}})$, with $y_{\NP} = \sfrac{\mu_{0} I_{\text{w}}}{2\pi B_{\text{r}}}$.
In the vicinity of the null point, the magnetic field components are approximately linear functions of the Cartesian coordinates.
Hence, the magnetic field can be approximated as a quadrupole magnetic field around the null point \cite{maj1932,fris1933}.
Along the atomic beam path, the approximate quadrupole magnetic field is \cite{wan2022,titgk2022} 
\begin{equation}\label{eq:Bquad}
    \vb{\bm{B}}_{\text{q}} = \frac{2\pi B_\text{r}^2 }{\mu_0 I_\text{w} } z_\text{a} \,\vuy + \frac{2\pi B_\text{r}^2 }{\mu_0 I_\text{w} }(y-y_{\text{\tiny{NP}}}) \, \vuz  \ .
\end{equation}
For the study of the time evolution of the atom inside the IR chamber both of the fields, $\vb{B}_{\text{exact}}$ and $\vb{B}_{\text{q}}$, are considered below.

After the IR chamber, a slit transmits one branch of electron spins initially polarized by SG1 and blocks the other branch.
The slit was positioned after the intermediate stage to obtain a sharper cut-off \cite{fris1933}.
In the forthcoming theoretical model, we track only the top transmitted branch with spin down, $m_{\subs}=-\sfrac{1}{2}$, at the entrance of the IR chamber and ignore the blocked branch.
However, the opposite choice of $m_{\subs}=+\sfrac{1}{2}$ yields exactly the same results in this model. 
The atoms that reach SG2 are further spatially split into two branches corresponding to the electron spin state with respect to the magnetic field direction. 
The final distribution of atoms is measured by scanning a hot wire along the $z$ axis while monitored by the microscope.
The probability of flip is then measured at different values of the electric current $I_{\text{w}}$.

\section{Theoretical description}\label{FStheo}

\begin{figure*}[ht]
\centering
    % \subfloat[\label{fig:ModelSchematic2}]
    \includegraphics[scale=0.9]{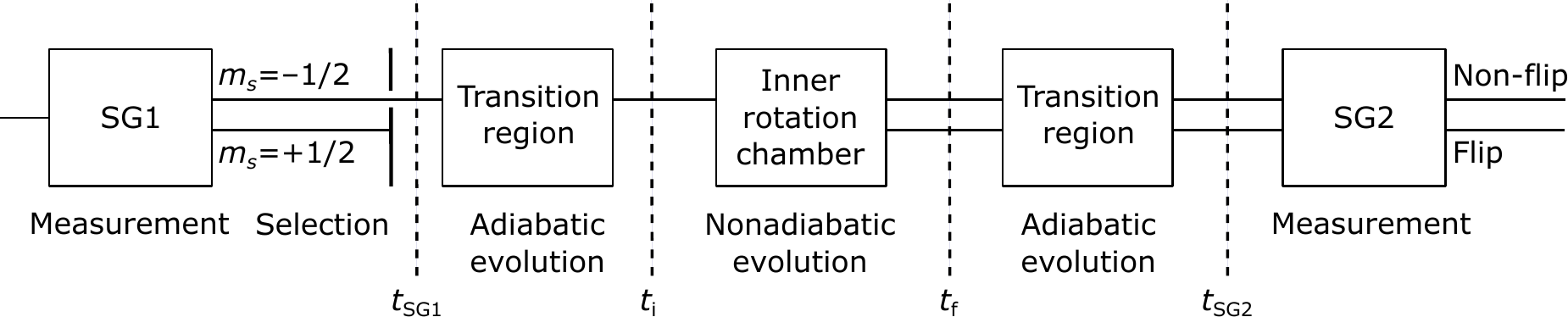}
    \caption{Schematic of the model considered in this study. The system consists of two measurements with SG1 and SG2. The inner rotation chamber that induces nonadiabatic transitions is modeled with the time-dependent von Neumann equation. The evolution from the end of SG1 and the filter to the entrance of the rotation chamber is modeled as an adiabatic evolution. Similarly, the evolution from the exit of the inner rotation chamber to the entrance of SG2 is modeled as adiabatic evolution.}
    \label{fig:model}
\end{figure*}

The time evolution of the noninteracting atoms in the beam traveling through the IR chamber of the Frisch--Segr\`e experiment is studied using standard quantum mechanics.
The whole setup is modeled in multiple stages as illustrated in Figure \ref{fig:model}. 
First, the output of SG1 and the slit is modeled as a pure eigenstate of the electron spin measurement in the $z$ direction. 
Since the gradient of the strong field in SG1 is not high enough, nuclear spin eigenstates do not separate during the flight. 
Hence, the nuclear state is assumed to be unaffected by SG1 and the slit. 
During the flight from SG1 to the entrance of the IR chamber, the state is assumed to vary adiabatically as in Figure \ref{fig:model}.
The fields in the transition regions were not reported; but when the IR chamber was turned off, $I_{\mathrm{w}}=\SI{0}{\ampere}$, no flipping was observed after SG2 \cite{fris1933ital}. 
Therefore, it can be assumed that outside the IR chamber, the state evolves adiabatically.
Later, the atom enters the IR chamber designed to induce nonadiabatic transitions. 
The evolution of the state in the IR chamber is modeled using the von Neumann equation, which is solved using numerical methods.
During the flight from the exit of the IR chamber to SG2, the state is assumed to vary adiabatically as in Figure \ref{fig:model}.
Finally, SG2 measures the probabilities in different electron spin eigenstates in the $z$ direction according to the Born principle. 

The density operator formalism is used for its capability to represent mixed states in quantum systems, offering a more complete description than pure states alone \cite{von1932,von2018}.
The time evolution of the density operator $\hat\rho$ is governed by the von Neumann equation \cite{von1932,von2018,alif2001,ben2009}:
\begin{equation}\label{eq:vonNeumann}
    \pdv{\hat{\rho}(t)}{t} = \frac{1}{i\hbar} \comm*{\hat{H}(t)}{\hat\rho(t)} \, ,
\end{equation}
where $\hat H(t)$ is the Hamiltonian of the system and $\hbar$ is the reduced Planck constant.
For the time-dependent Hamiltonian $\hat{H}(t)$, we introduce the instantaneous eigenstates $\ket{\psi_{j}(t)}$ and eigenenergies $\mathcal{E}_{j}(t)$ such that
\begin{equation}
    \hat{H}(t) \ket{\psi_{j}(t)} = \mathcal{E}_{j}(t) \ket{\psi_{j}(t)} ,
\end{equation}
where $j$ can take a finite number of values for the spin system considered here.
In the basis of the instantaneous eigenstates of the Hamiltonian, from \eqref{eq:vonNeumann} the matrix elements of the density operator, $\rho_{j,k}(t) = \matrixel{\psi_{j}(t)}{\hat{\rho}(t)}{\psi_{k}(t)}$, evolve according to
\begin{widetext}
\begin{multline}\label{eq:rhot}
    \pdv{\rho_{j,k}(t)}{t} = 
    \Biggl[ \frac{1}{i\hbar}\left(\mathcal{E}_{j}(t) -\mathcal{E}_{k}(t)\right) - \bra{\psi_{j}(t)} \pdv{\ket{\psi_{j}(t)}}{t} + \bra{\psi_{k}(t)} \pdv{\ket{\psi_{k}(t)}}{t} \Biggr] \, \rho_{j,k}(t)\\
    + \sum_{r\neq q} \frac{\matrixel{\psi_{l}(t)}{\pdv{\hat{H}(t)}{t}}{\psi_{k}(t)}}{\mathcal{E}_{k}(t)-\mathcal{E}_{l}(t)} \, \rho_{j,l}(t) - \sum_{r\neq p} \frac{\matrixel{\psi_{j}(t)}{\pdv{\hat{H}(t)}{t}}{\psi_{l}(t)}}{\mathcal{E}_{l}(t)-\mathcal{E}_{k}(t)} \, \rho_{l,k}(t) \ .
\end{multline}   
\end{widetext}

In particular, the elements in the diagonal $\rho_{j,j}(t)$,  corresponding to the probabilities of finding the quantum system in the eigenstate of the Hamiltonian, follow
\begin{equation}
        \pdv{\rho_{j,j}(t)}{t} = 2\Re\left[ \sum_{r\neq p} \frac{\matrixel{\psi_{l}(t)}{\pdv{\hat{H}(t)}{t}}{\psi_{j}(t)}}{\mathcal{E}_{j}(t)-\mathcal{E}_{l}(t)} \, \rho_{j,l}(t) \right] .
\end{equation}
In the adiabatic approximation, since the time derivative of the Hamiltonian is small compared to the energy difference, we set \cite{agunp2000,griffiths2018introduction}
\begin{equation} \label{eq:adiabaticapprox}
    \frac{\matrixel{\psi_{l}(t)}{\pdv{\hat{H}(t)}{t}}{\psi_{j}(t)}}{\mathcal{E}_{j}(t)-\mathcal{E}_{l}(t)}  \rightarrow 0 .
\end{equation} 
Therefore, for the adiabatic evolution, the populations in the instantaneous eigenstates do not change over time
\begin{equation}\label{eq:timeinvariance}
    \pdv{\rho_{j,j}(t)}{t} = 0 \, .
\end{equation}
If the system's Hamiltonian changes quickly relative to the energy gap, the above approximation fails, leading to nonadiabatic transitions.

Let us consider the quantum system for a neutral alkali atom, composed of the spin $S=\sfrac{1}{2}$ of the valence electron and the spin $I$ of the nucleus.
In an external magnetic field $\vb{B}$, the electron Zeeman term $\hat{H}_{\mathrm{e}}$ describes the interaction between the electron magnetic moment and the field via \cite{griffiths2018introduction}
\begin{equation}\label{eq:He1}
    \hat H_{\mathrm{e}}= -\opmue \cdot \vb{\bm{B}} \, ,
\end{equation}
where $\opmue$ is the quantum operator for the electron magnetic moment.
Furthermore, $\opmue = \gamma_{\mathrm{e}} \hat{\vb{S}}$, where $\gamma_{\mathrm{e}}$ denotes the gyromagnetic ratio of the electron; the electron spin operator $\hat{\vb{S}}= \frac{\hbar}{2} \hat{\vb{\bm{\sigma}}}$, with the Pauli vector $\hat{\vb{\bm{\sigma}}}$ consisting of the Pauli matrices $\left\{ \sigma_{x}, \sigma_{y}, \sigma_{z}  \right\}$.
% $\gamma_{\mathrm{e}} = \SI{-1.760 859 630 23(53)e11}{\radian\per\second\per\tesla}$
Substitutions yield
\begin{equation}\label{eq:He2}
    \hat{H}_{\mathrm{e}} = -\gamma_{\mathrm{e}} \frac{\hbar}{2} \, \hat{\vb{\bm{\sigma}}} \cdot  \vb{B} \, .
\end{equation}
The $(2S+1)$-dimensional Hilbert space $\mathcal{H}_{\mathrm{e}} = \mathrm{span}\left( \ket{S,m_{\subs}} \right)$, with $m_{\subs}=-S,\dots,S$, and $\ket{S,m_{\subs}}$ being the eigenvectors of $\hat{S}_{z}$ . 

The nuclear Zeeman Hamiltonian $\hat{H}_{\mathrm{n}}$ describes the interaction of the nuclear magnetic moment with the external magnetic field:
\begin{equation}\label{eq:Hn1}
    \hat H_{\mathrm{n}}= -\opmun \cdot \vb{\bm{B}} \, ,
\end{equation}
where $\opmun=\gamma_\mathrm{n} \hat{\vb{I}}$ denotes the quantum operator for the nuclear magnetic moment, $\gamma_\mathrm{n}$ the nuclear gyromagnetic ratio for the atomic specie, and $\hat{\vb{I}}$ the nuclear spin quantum operator for spin $I$.
Therefore, we can write $\hat{\vb{I}}= \frac{\hbar}{2} \hat{\vb{\bm{\tau}}}$, with $\hat{\vb{\bm{\tau}}}$ being the generalized Pauli vector constructed with the generalized Pauli matrices of dimension $2I+1$, namely $\left\{  \tau_{x}, \tau_{y}, \tau_{z}\right\}$, satisfying $\comm{\tau_{j}}{\tau_{k}}=2i\epsilon_{jkl}\tau_{l}$.
Substitutions produce
\begin{equation}\label{eq:Hn2}
    \hat H_{\mathrm{n}}= -\gamma_{\mathrm{n}}\frac{\hbar}{2} \,  \hat{\vb{\bm{\tau}}} \cdot \vb{B} \, .
\end{equation}
A basis for the $(2I+1)$-dimensional Hilbert space $\mathcal{H}_{\mathrm{n}}$ can be obtained from the eigenvectors of $\hat{I}_{z}$, such that $\mathcal{H}_{\mathrm{n}}=\mathrm{span}\left( \ket{I,m_{\subi}} \right)$ with $m_{\subi}=-I,\dots,I$. 

The interaction between the magnetic dipole moments of the nucleus and the electron gives the hyperfine structure (HFS) term $\hat{H}_{\HFS}$.
In terms of the electron and nuclear spin operators, the Hamiltonian is written as
\begin{equation}\label{eq:Hint1}
    \hat{H}_{\HFS} = \frac{2\pi a_{\HFS}}{\hbar}  \  \hat{\vb{\bm{I}}} \cdot \hat{\vb{\bm{S}}} \ ,
\end{equation}
where $a_{\HFS}$ reflects the coupling strength.

Then, the total Hamiltonian of the combined system consisting of the electron and nuclear spins under an external magnetic field is
\begin{equation}\label{eq:fullH}
    \hat H_{\mathrm{total}} = \hat H_{\mathrm{e}} + \hat H_{\mathrm{n}} + \hat H_{\HFS} \, .
\end{equation}
The $(2S+1)(2I+1)$-dimensional Hilbert space for the combined nuclear--electron spin system is $\mathcal{H} = \mathcal{H}_{\mathrm{n}}\otimes\mathcal{H}_{\mathrm{e}}=\mathrm{span}\left( \ket{m_{\subi},m_{\subs}} \right)$; where for simplicity of notation we have dropped the $S$ and $I$ labels.

The Frisch--Segrè experiment used $^{39}\mathrm{K}$; for this isotope, the nuclear spin is $I=\sfrac{3}{2}$, the nuclear gyromagnetic ratio is $\gamma_{\mathrm{n}} = \SI{1.2500612(3)e7}{\radian\per\second\per\tesla}$, and the experimentally measured hyperfine constant is $a_{\text{exp}}=\SI{230.8598601(3)}{\mega\hertz}$  \cite{ariiv1977}.
The terms of the nuclear--electron spin Hamiltonian $\hat H_{\mathrm{total}}$ in \eqref{eq:fullH} are explicitly expressed as \cite{lev2008, schmied2020using}
\begin{equation}\label{eq:He}
\begin{aligned}
    \hat{H}_{\mathrm{e}} &= - \gamma_\mathrm{e} \frac{\hbar}{2} \, \hat\tau_0 \otimes (B_x \hat\sigma_x + B_y \hat\sigma_y + B_z \hat\sigma_z ) \\
    &= -\gamma_\mathrm{e} \frac{\hbar}{2} \ \hat\tau_0 \otimes 
    \pmqty{ B_z & B_x-i B_y \\ B_x+i B_y & -B_z } \ ,
\end{aligned}
\end{equation}

\begin{widetext}
\begin{multline}\label{eq:Hn}
    \hat{H}_{\mathrm{n}} = -\gamma_\mathrm{n}  \frac{\hbar}{2}  \, (B_x \hat\tau_x + B_y \hat\tau_y + B_z \hat\tau_z ) \otimes \hat\sigma_0 \\
    = -\gamma_\mathrm{n}  \frac{\hbar}{2} \,  
    \pmqty{3B_z & \sqrt{3}(B_x-iB_y) & 0 & 0 \\ \sqrt{3}(B_x+iB_y) & B_z & 2(B_x-iB_y) & 0 \\ 0 & 2(B_x+iB_y) & -B_z & \sqrt{3}(B_x-iB_y) \\ 0 & 0 & \sqrt{3}(B_x+iB_y) & -3B_z } 
   \otimes \hat\sigma_0 \ ,
\end{multline}
\end{widetext}

\begin{equation}\label{eq:Hint}
\begin{aligned}
    \hat{H}_{\mathrm{\HFS}} &= \frac{\pi}{2} \hbar \, a_{\mathrm{\HFS}}  \left(\hat\tau_x \otimes \hat\sigma_x + \hat\tau_y \otimes \hat\sigma_y +\hat\tau_z \otimes \hat\sigma_z \right) \\
    &= \frac{\pi}{2} \hbar \, a_{\text{\HFS}}
    \pmqty{
    3 & 0& 0& 0& 0& 0& 0& 0\\
    0& -3& 2\sqrt{3}& 0& 0& 0& 0& 0\\
    0& 2\sqrt{3}& 1& 0& 0& 0& 0& 0\\
    0& 0& 0& -1& 4& 0& 0& 0\\
    0& 0& 0& 4& -1& 0& 0& 0\\
    0& 0& 0& 0& 0& 1& 2\sqrt{3}& 0\\
    0& 0& 0& 0& 0& 2\sqrt{3}& -3& 0\\
    0& 0& 0& 0& 0& 0& 0& 3 } \ ,
\end{aligned}
\end{equation}
in the $\left\{ \ket{m_{\subi},m_{\subs}}\right\}$ basis, where $\hat\sigma_0$ and $\hat\tau_0$ are the 2-dimensional and 4-dimensional identity matrices, respectively.
This Hamiltonian has been validated numerically by comparing the eigenvalues with the solutions from the exact Breit--Rabi formula \cite{brer1932} with respect to the external field. 

\subsection{Adiabatic evolution}

As depicted in Figure \ref{fig:model}, the system undergoes an adiabatic evolution from SG1 (polarizing magnet) to the entrance of the rotation chamber, $t\in\left[ t_{\SGone},t_{\mathrm{i}} \right]$, as well as from the exit of the inner rotation chamber to SG2 (the analyzing magnet), $t\in\left[ t_{\mathrm{f}},t_{\SGtwo}\right]$. 
From \eqref{eq:adiabaticapprox}, we have
\begin{align}
    \rho_{j,j}(t_{\SGone})&=\rho_{j,j}(t_{\mathrm{i}}) , & \rho_{j,j}(t_{\SGtwo})&=\rho_{j,j}(t_{\mathrm{f}}).
\end{align}

The quantum state of the atoms after SG1 and the filter is a pure state for the electron but maximally mixed for the nuclear spin \cite{rab1936}. 
Hence, the density matrix is diagonal following
\begin{equation}\label{eq:rhoSG1psi}
    \hat{\rho}(t_{\SGone}) = \sum_{j}^{} \rho_{j,j}^{\text{initial}} \ketbra{\psi_{j}(t_{\SGone})} \ .
\end{equation}

The measurement probabilities at SG2 can be directly obtained from the state at $t_{\mathrm{f}}$ from
\begin{equation}\label{eq:pSG2}
    p_{j} = \matrixel{\psi_{j}(t_{\text{f}})}{\hat{\rho}(t_{\text{f}})}{\psi_{j}(t_{\text{f}})}\ .
\end{equation}

\subsection{Nonadiabatic evolution}

In the IR chamber, the external magnetic field is not homogeneous; instead, along the beam path the magnetic field rapidly changes its direction and magnitude.
The IR chamber of the FS experiment was specially designed to induce nonadiabatic variations of the magnetic field \cite{phis1932,fris1933}.
For such behavior the field has to be sufficiently weak and the variation of its direction sufficiently fast, such that the frequency of rotation of the magnetic field is large compared to the Larmor frequency \cite{maj1932, Bassani2006}.
The conditions for nonadiabatic rotations are satisfied near $y=y_{\NP}$ along the beam path \cite{titgk2022}.

An exact closed-form analytical time-dependent solution for the density operator in the IR chamber has not been obtained.
To calculate a numerical solution, we discretize the von Neumann equation \eqref{eq:vonNeumann}.
We used several different differential equation solvers to validate the solutions \cite{bezek2017,racn2017}, one of which is the second-order Runge--Kutta method \cite{suli2003introduction}:
\begin{subequations}
\begin{align} \label{eq:H0_RK}
    \hat\rho(t+\frac{\Delta t}{2}) &= \hat\rho(t) -  \frac{\Delta t}{2} \frac{i}{\hbar} \comm{\hat H(t)}{\hat\rho(t)} \ , \\
    \hat\rho(t+\Delta t) &= \hat\rho(t) -  \Delta t \frac{i}{\hbar} \comm{\hat H(t+\frac{\Delta t}{2})}{\hat\rho(t+\frac{\Delta t}{2})} \ ,
\end{align} 
\end{subequations}
where $\Delta t$ is the temporal step size.

\section{Results}

Historically, the first attempt to describe nonadiabatic rotations was made by Majorana with a model involving only the electron Zeeman term \cite{maj1932}. 
In Section \ref{excludinghfs}, the same model is solved numerically while ignoring any nuclear effect. 
Improving on Majorana's solution, Rabi considered the effect of the nuclear spin \cite{rab1936}.
Here, we explore the same model numerically in Section \ref{includinghfs}. 
Our results indicate that in the IR chamber, the field strength is weak enough that the effect of the nuclear spin cannot be neglected. 

\subsection{Excluding hyperfine interaction}\label{excludinghfs}

We first consider the case when the Hamiltonian $\hat H$ is $\hat H_{\text{e}}$ in \eqref{eq:He2}, excluding the nuclear Zeeman and hyperfine effects.
The analytical asymptotic solution for this model was found using the quadrupole field approximation by Majorana \cite{maj1932} and applied to the Frisch--Segr\`e experiment \cite{fris1933}. 
The flip probability is given by the well-known Landau-Zener-Stuckelberg-Majorana (LZSM) model:
\begin{equation}\label{eq:pLZSM}
p_{\text{LZSM}}=\sin^2(\alpha/2)=\exp(-\frac{\pi}{2} k)    ,
\end{equation}
where the adiabaticity parameter is defined as (see Appendix \ref{app:Badd} for details) \cite{maj1932,ivasn2023}
\begin{equation}\label{eq:k}
     k = \frac{|\gamma_{\mathrm{e}}|}{v} \frac{2\pi B_{r}^{2}}{\mu_{0} I_\text{w}} z_{\mathrm{a}}^2 \, .
\end{equation}

\begin{figure}
    \centering
    \includegraphics[width=0.95\linewidth]{ 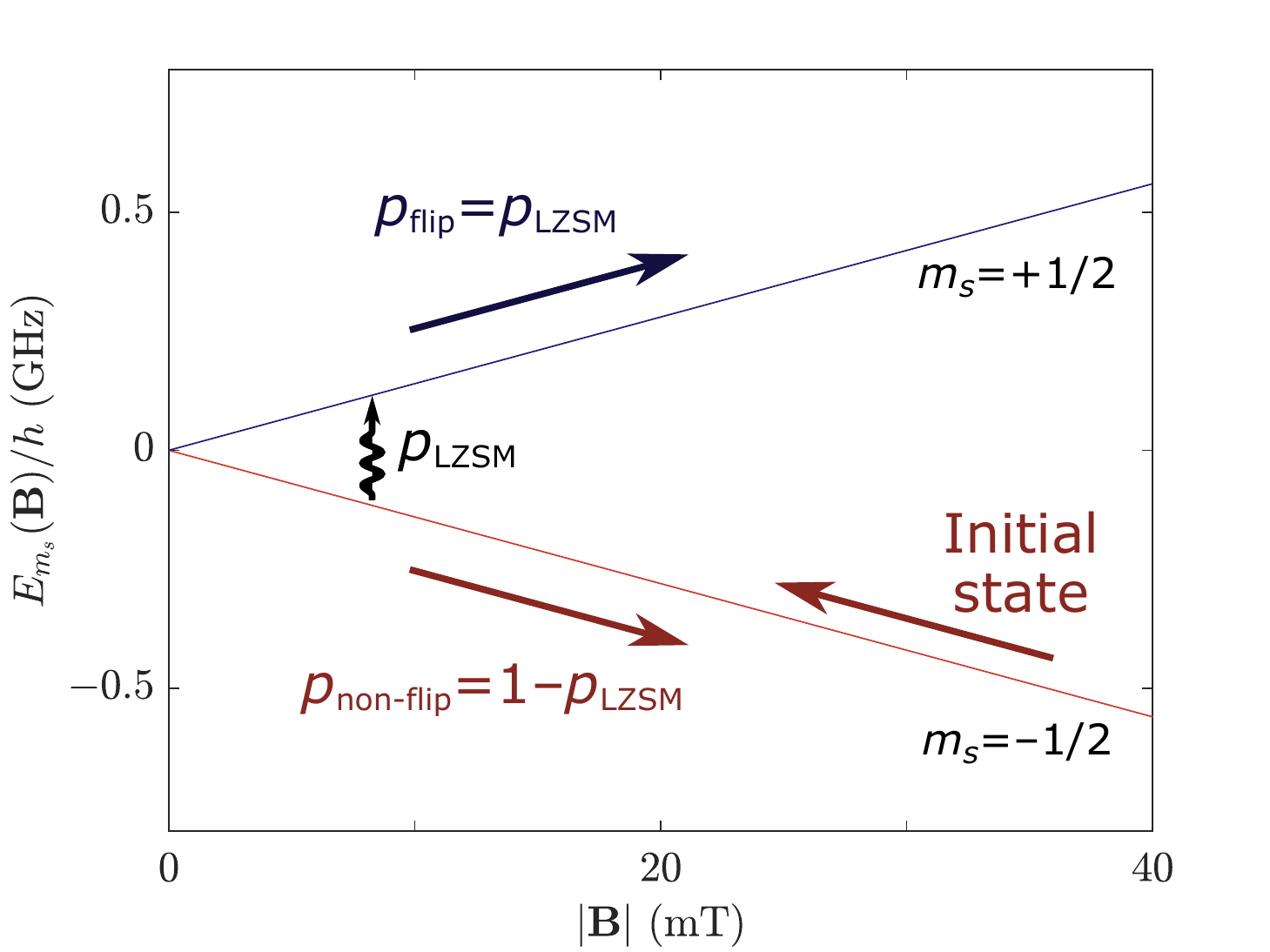}
    \caption{
    Eigenenergy-based visualization of the nonadiabatic transition of the electron spin. 
    The initial electron spin is in the $m_{\subs}=-1/2$ state. 
    As the atom nears the null point, the field strength reduces. 
    When eigenenergies converge, the rapid field rotation triggers a nonadiabatic transition. 
    In the quadrupole field approximation, the nonadiabatic transition can be described through the Landau-Zener-Stückelberg-Majorana (LZSM) model \cite{maj1932, ivasn2023}.
    }
    \label{fig:Zeemane}
\end{figure}

Here, we numerically solve this model for both the exact and quadrupole fields. 
In modeling adiabatic evolution as discussed in Section \ref{FStheo}, we introduce an instantaneous eigenstate $\ket{\psi_{m_{\subs}}(t)}$ with the associated instantaneous eigenenergy $\mathcal{E}_{m_{\subs}}(t)= m_{\subs} \gamma_{\mathrm{e}}\hbar \vert \vec{B}(t) \vert$.
As the atom nears the null point, the instantaneous eigenenergies become asymptotically degenerate, and the state transitions nonadiabatically between the instantaneous eigenstates as visualized in Figure \ref{fig:Zeemane}.

\begin{figure}[!t]
\captionsetup[subfloat]{justification=Justified, singlelinecheck=false}
\centering
    \subfloat[\label{fig:timetrace_electron}]{\includegraphics[width=0.95\columnwidth, trim={0 0 15 12}, clip]{ 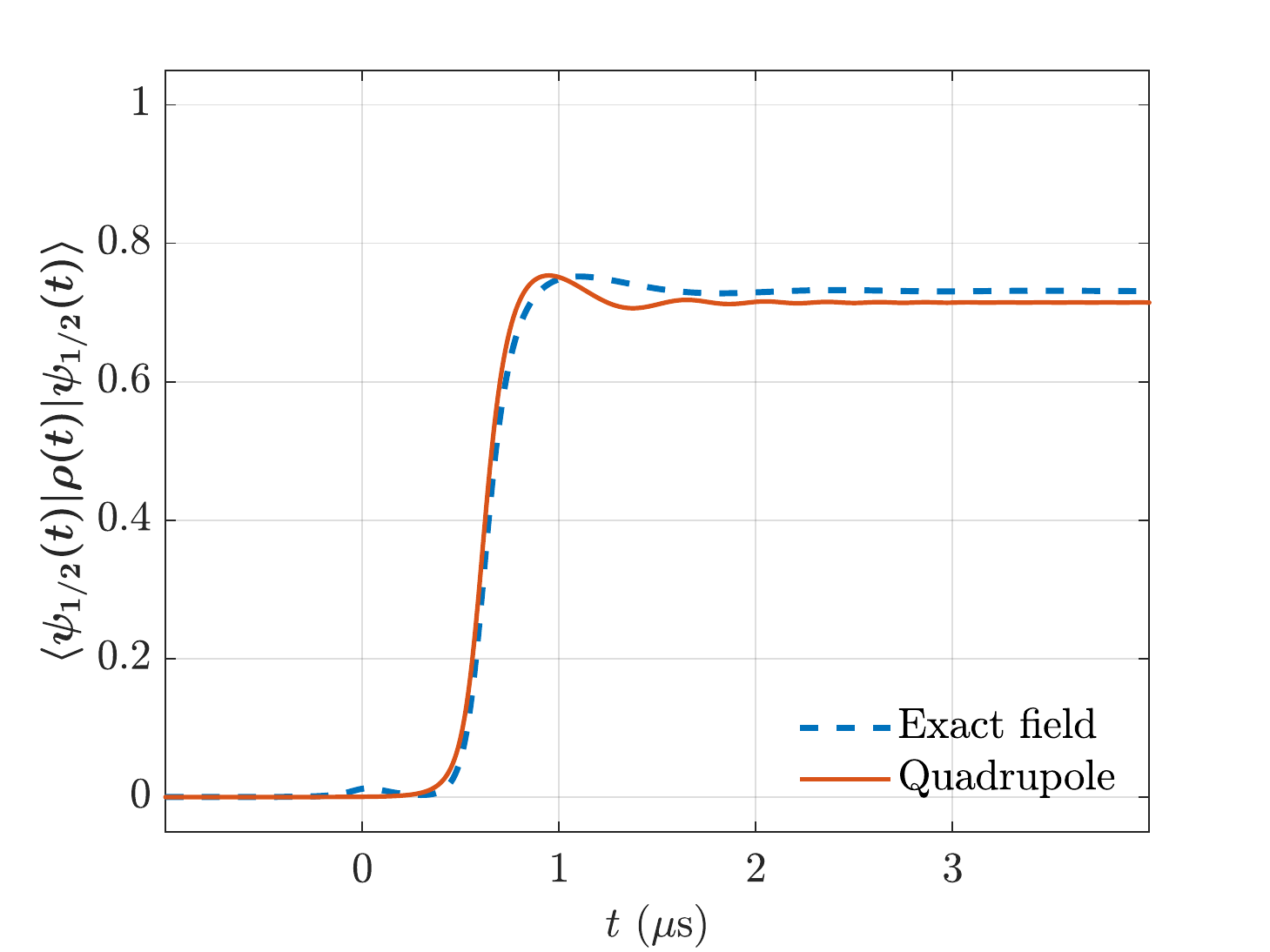}}
    \\
    \subfloat[\label{fig:curve_electron}]{\includegraphics[width=0.95\columnwidth, trim={0 0 15 12}, clip]{ 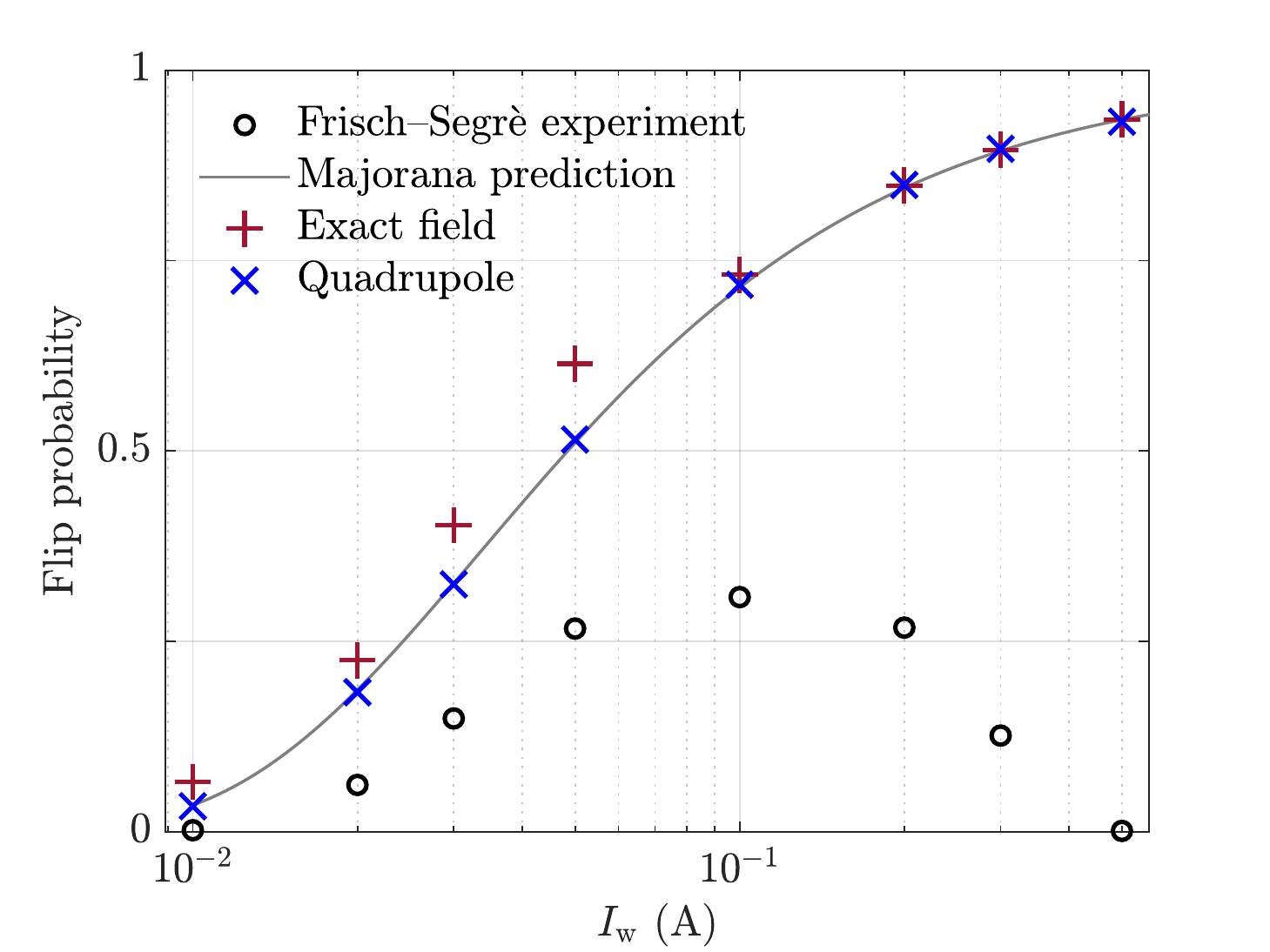}}
    \centering
    
    \caption{ Electron spin only model, where hyperfine interaction is ignored. (a) Time evolution of $\matrixel{\psi_{\sfrac{1}{2}}(t)}{\hat{\rho}(t)}{\psi_{\sfrac{1}{2}}(t)}$ for the  exact and quadrupole fields at the wire current $I_{\text{w}}=\SI{0.1}{\ampere}$. (b) Flip probability of the electron spin versus the wire current. The numerical simulations match Majorana's prediction \cite{maj1932} but not the experimental observation \cite{fris1933}.  }
    \label{fig:electronresults}
\end{figure}

Figure \ref{fig:timetrace_electron} shows the evolution of $\matrixel{\psi_{\sfrac{1}{2}}(t)}{\hat{\rho}(t)}{\psi_{\sfrac{1}{2}}(t)}$ over the flight of the atom in the IR chamber, where $I_{\text{w}}=\SI{0.1}{\ampere}$. 
The evolution based on the quadrupole field approximation closely follows that based on the exact field, indicating the accuracy of the field approximation.

Figure \ref{fig:curve_electron} shows the flip probability of the electron spin observed in SG2 as $\ket{\psi_{\sfrac{1}{2}}}$ for the exact and quadrupole fields at different wire currents. 
The numerical prediction using the quadrupole approximation agrees exactly with Majorana's analytical prediction \cite{maj1932} and closely with the numerical prediction using the exact field. 
The coefficients of determination $R^2$ between the numerical predictions and the experimental data are, however, $-18.9$ and $-19.9$ for the exact and quadrupole fields, respectively.
Therefore, this model does not predict the experimental observations.

\subsection{Including hyperfine interaction}\label{includinghfs}

We now generalize the Hamiltonian $\hat H$ to $\hat H_{\text{total}}$ in \eqref{eq:fullH} by including both the nuclear Zeeman and hyperfine effects. 
The total spin for the system assumes the values of $F=I\pm\frac{1}{2}$.
Let the instantaneous eigenstate $\ket{\psi_{j}(t)} $ be $ \ket{\psi_{m_{\subi},m_{\subs}}(t)}$. 

\begin{figure}
    \centering
    \includegraphics[width=0.95\linewidth]{ 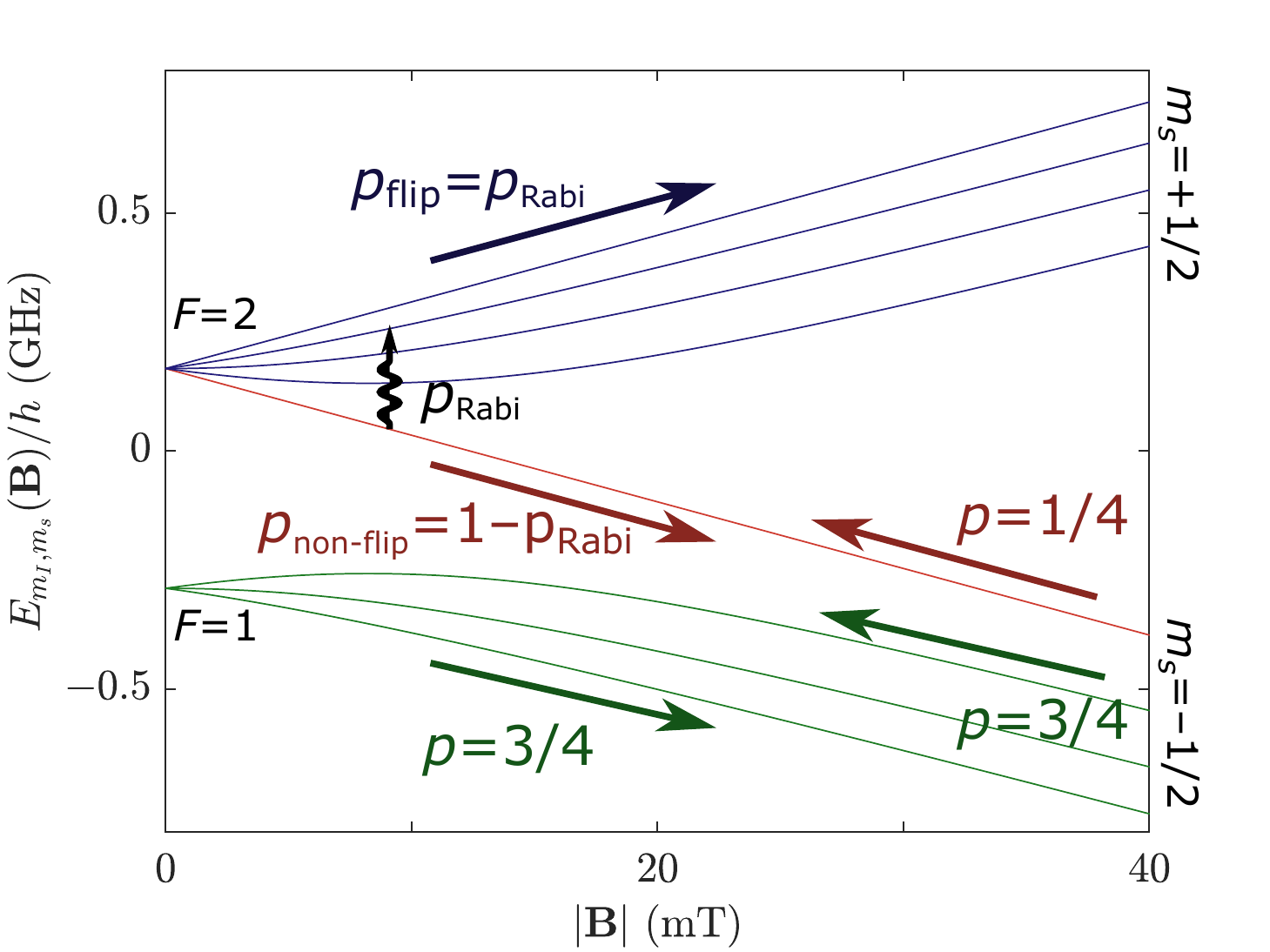}
    \caption{Eigenenergy-based visualization of the nonadiabatic transition of the nucleus and electron spins. The initial state is in a mixture of $m_{\subs}=-1/2$ states. As the atom approaches the null point, the state mostly adiabatically transitions into low-field eigenstates. Near the null point, when the $F=2$ states have similar eigenenergies, nonadiabatic transitions occur. As the atom leaves the null point region, the state again follows adiabatically to the high-field eigenstates. However, a portion of the population reaches the $m_{\subs}=+1/2$ states due to the nonadiabatic transition. When the quadrupole field is used, the nonadiabatic transition can be modeled by Rabi's derivation \cite{rab1936}.}
    \label{fig:Zeeman}
\end{figure}

Building on Majorana's work, Rabi developed an asymptotic solution that incorporates the influence of the nuclear spin \cite{rab1936}.
Rabi's approach has been visualized in Figure \ref{fig:Zeeman}. 
Pairs of eigenstates with different $F$ values are too far apart in eigenenergies to nonadiabatically transition to each other. 
In contrast, the eigenstates with the same $F$ values become asymptotically degenerate as the atom approaches the null point.
Therefore, eigenstates within each $F$ manifold nonadiabatically couple to each other, and the approximation in \eqref{eq:adiabaticapprox} no longer holds.
Rabi solved the nonadiabatic transition in each $F$ manifold as a single rotation in a $(2F+1)$-dimensional Hilbert space. 
The flip probability between individual eigenstates ($m_F\rightarrow m_F'$) after a rotation can be calculated using the Wigner-d matrix \cite{rab1936, Torrey1937}:
\begin{equation}\label{eq:wigner}
    p_{m_F,m_F'}^F(\alpha) = |d_{m_F,m_F'}^F(\alpha)|^2 ,
\end{equation}
where the angle of rotation can be found through 
\begin{equation}\label{eq:rabi_0}
\sin^2(\alpha/2)=e^{-\frac{\pi}{2} k'} .
\end{equation}
The adiabaticity parameter $k'$ is approximated as 
\begin{equation}\label{eq:k_rabidelta}
 k' \approx \frac{1}{2I+1} k  \, ,
\end{equation}
which accounts for the modified Land\'e $g$-factor. 

The total transition probability between the $m_F=-F$ state and the $m_F \neq -F$ states is found with a summation:
\begin{equation}\label{eq:rabi}
    p_{\rabi} = \frac{1}{2I+1} \sum_{m_F\neq-F} p_{m_F,-F}^F \ ,
\end{equation}
where the prefactor denotes the initial populations, which are assumed to be equal. 
The maximally mixed initial state is given by
\begin{equation}\label{eq:maximimallymixed}
    \hat{\rho}(t_{\text{i}}) = \frac{1}{2I+1} \sum_{m_{\subi}}^{} \ketbra{\psi_{{m_{\subi},-\sfrac{1}{2}}}(t_{\text{i}})} \ .
\end{equation}

\begin{figure}[!t]
\captionsetup[subfloat]{justification=Justified, singlelinecheck=false}
\centering
        \subfloat[\label{fig:qm_timetrace_adiabatic}]{\includegraphics[width=0.95\columnwidth]{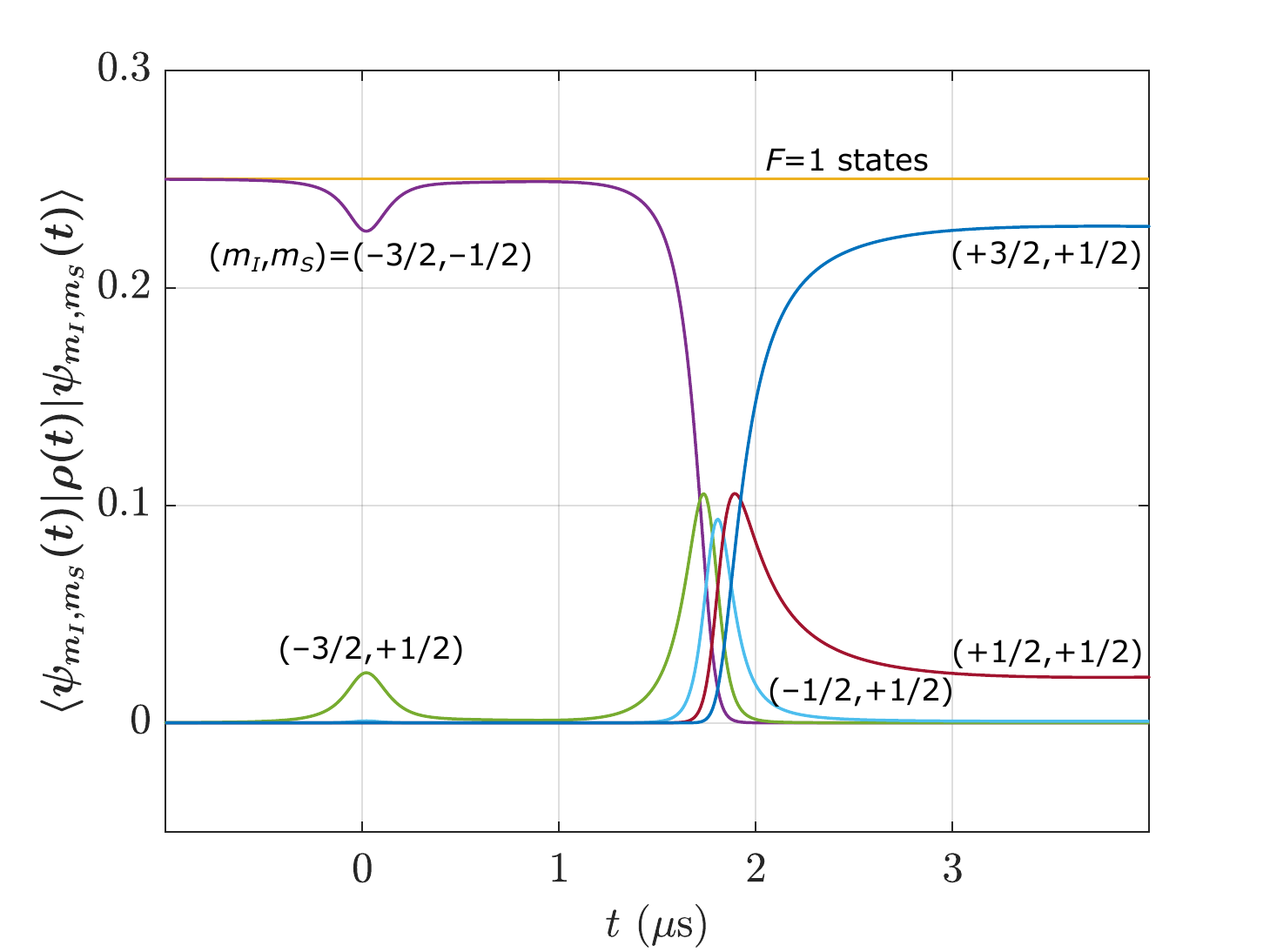}}
    \\
    \subfloat[\label{fig:curve_qm_adiabatic}]{\includegraphics[width=0.95\columnwidth]{ 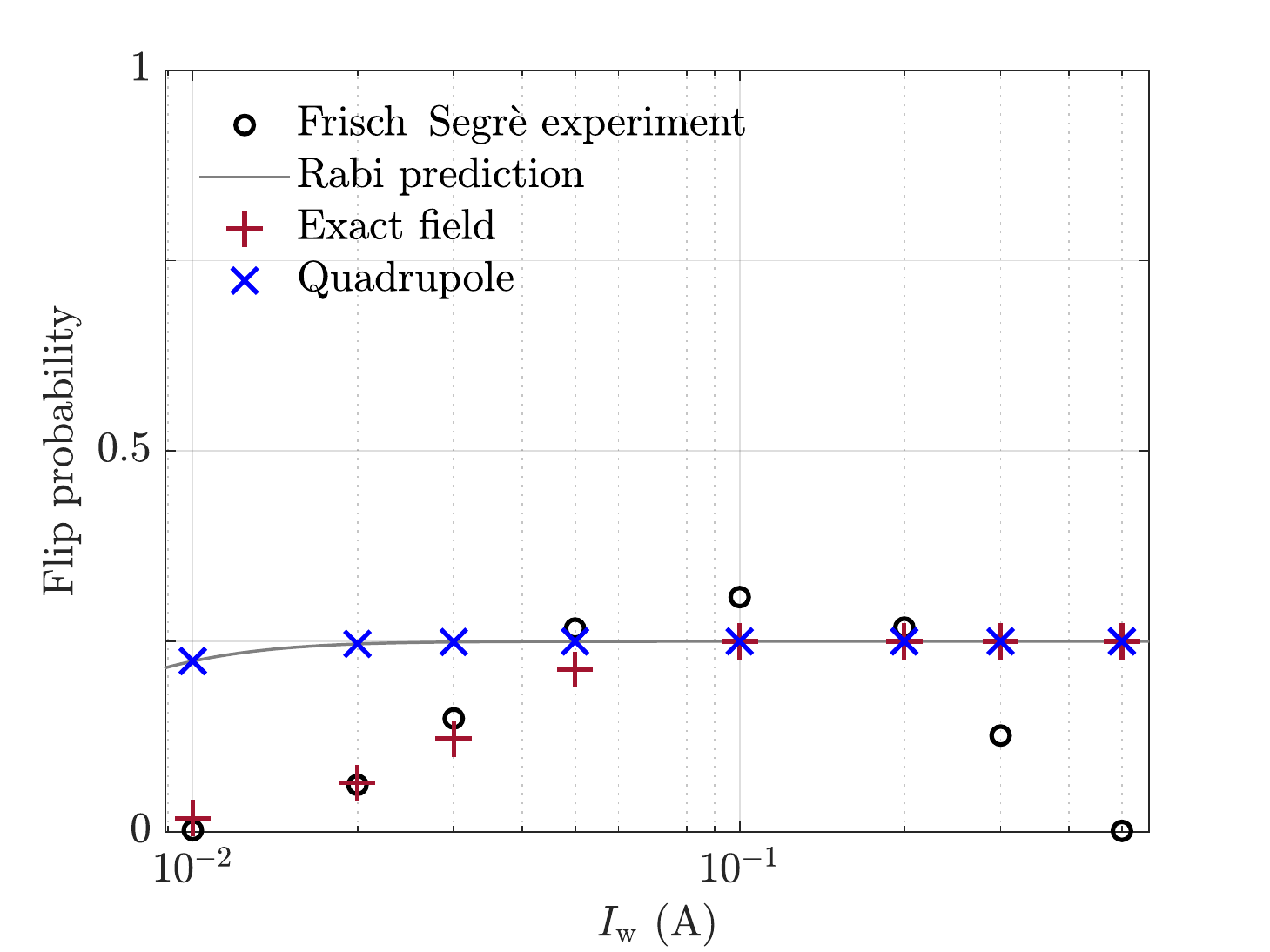}}
    \centering
         
    \caption{Nucleus-electron spin model, where the nuclear effects are included. (a) Evolution of the populations for the exact field within the IR chamber at $I_{\text{w}}=\SI{0.3}{\ampere}$. Near the null point, the populations exchange between $\ket{\psi_{-3/2,-1/2}(t)}$ and states with $m_{\subs} = +1/2$. Far from the null point, the evolution of the state is mostly adiabatic. (b) Flip probability of the electron spin for the exact and quadrupole fields. The numerical predictions do not match the experimental observation well.    }
    \label{fig:curve_adiabatic}
\end{figure}

Figure \ref{fig:qm_timetrace_adiabatic} illustrates the evolution of the populations,  $\matrixel{\psi_{m_{\subi},m_{\subs}}(t)}{\hat{\rho}(t)}{\psi_{m_{\subi},m_{\subs}}(t)}$, around the null point. 
The state follows the instantaneous eigenstates adiabatically during the majority of the flight. 
Above the wire, there is a small nonadiabatic evolution that recovers. 
Hence, the majority of the nonadiabatic transition occurs around the null point. 
Apart from the behavior immediately above the wire, the assumption that the state follows the instantaneous eigenstates outside the null point is accurate. 
For a lower current, where the null point is closer to the wire, the behavior right above the wire affects the transition near the null point.
Therefore, the quadrupole approximation for the low currents is inaccurate (to be shown below). 

Figure \ref{fig:curve_qm_adiabatic} shows the flip probabilities predicted by the numerical solution in comparison to Rabi's analytical solution \cite{rab1936} and the experimental observation \cite{fris1933}.
The inaccuracy of the quadrupole approximation explains the discrepancy between the numerical solutions in Figure \ref{fig:curve_qm_adiabatic}.
Furthermore, the numerical solution with the quadrupole approximation matches Rabi's solution as expected. 
The coefficients of determination $R^2$ of our model for the exact and quadrupole fields in relation to the experimental observation are $0.19$ and $-0.67$, respectively.
Our standard quantum mechanical model or Rabi's solution, even if hyperfine interaction is considered, does not predict the experimental observation well in the high current regime.

\section{Discussion}\label{discussion}

As a natural extension of the first SG experiment, FS aimed to implement cascaded quantum measurements using two SG apparatuses with a nonadiabatic spin flipper in between. 
Since the SG apparatuses here cannot distinguish nuclear eigenstates \cite{fris1933}, it was conceivable for Majorana to ignore the nuclear effect \cite{maj1932}.
However, as shown in Section \ref{excludinghfs}, such an electron-only model cannot predict the FS observations. 
Along the beam path, the atom enters a magnetically weak region where the effect of the nuclear spin is critical as shown in Section \ref{includinghfs}. 
While factoring in nuclear effects greatly enhances the theory-experiment match, it still fails to model the highly nonadiabatic regime. 
Regarding the decrease of the flip probability as a function of increasing current in the high-current regime, FS mentioned that it could be observed if there was a remnant field along the propagation axis \cite{fris1933ital}.
In Appendix \ref{app:Badd}, we tried to fit the observations with the remnant field as a free parameter disregarding the reported value.

Furthermore, the peak flip probability in the experimental data exceeds $\sfrac{1}{4}$ while the theoretical models here cannot. 
There are at least two possibilities for the theoretical model to exceed $\sfrac{1}{4}$. 
Either the initial nuclear state is not maximally mixed as in \eqref{eq:maximimallymixed} or the hyperfine interaction strength, $a_{\HFS}$, is orders of magnitude smaller. 
Surprisingly, recent ``semi-classical'' studies \cite{wan2022,titgk2022,hetk2022} have been able to predict the Frisch--Segr\`e observations without fitting.

In Appendix \ref{app:modrho}, we have considered various other initial nuclear states. 
Since SG1 cannot distinguish nuclear eigenstates, one might question how the nuclear state must be modeled after SG1 and the slit. 
In Appendix \ref{app:moda}, we consider different hyperfine interaction strengths. 
The hyperfine interaction term, first written by Fermi, uses an interaction strength, $a_{\text{HFS}}$, theoretically calculated via the Fermi contact interaction \cite{fer1930, Griffiths1982}. 
However, for any atom except the hydrogen isotopes, the theoretical values are orders of magnitude different than the spectroscopically measured values \cite{ariiv1977, Ramsey1956}.
While this discrepancy might be due to the ill-defined nature of the Fermi contact interaction \cite{Soliverez1980}, using such theoretical values to model the FS experiment can result in flip probabilities higher than $\sfrac{1}{4}$. 
Ultimately, none of the variants of standard quantum mechanical approaches in the appendices follow the reported experiment but have been explored to stimulate discussion.

\section{Conclusions}\label{concl}
Simulating the FS experiment \cite{fris1933, fris1933ital} using a standard quantum mechanical model has yielded the following conclusions: 
\begin{itemize}
  \item The FS observations cannot be replicated by modeling only the electron spin without hyperfine coupling (Section \ref{excludinghfs}). Considering hyperfine coupling significantly improves the predictions (Section \ref{includinghfs}). 
  \item The FS observations cannot be closely replicated by modeling the atom as a pair of electron and nuclear spins without modifying the reported experimental parameters (Appendix \ref{app:Badd}), the initial nuclear state (Appendix \ref{app:modrho}), or the hyperfine interaction coefficient (Appendix \ref{app:moda}).
\end{itemize}

Based on some of the non-standard variants that can improve the model prediction of the FS observation (see Appendices), one might question the following:
\begin{itemize}
  \item What is the nuclear spin state before and after each SG apparatus? How does the nuclear spin state affect the electron spin-flip?
  \item Do we need a more sophisticated model of the atom (especially, the nucleus) to understand and predict the SG and nonadiabatic FS experiments?
  \item Do the SG apparatuses in the FS experiment truly follow the Born principle? Is there a (hidden) variable, such as the nuclear spin, affecting the FS measurement? % (see Appendix \ref{app:squared})
\end{itemize}

Despite the prevalent understanding of multi-stage Stern-Gerlach experiments, our models here fall short of accurately explaining the initial experiments \cite{fris1933, fris1933ital}. 
Later multi-stage SG experiments include different designs of the SG apparatuses and spin flippers, and the associated models used free parameters for fitting \cite{hig1975, higr1978, sch1982, schb1983, macjf2013, mardz2021}.
Given the foundational importance of the multi-stage SG experiments as cascaded quantum measurements, we believe that the mismatch between the theory and the experiment merits further investigation.

\begin{acknowledgments}
We thank Xukun Lin for scrutinizing the source codes, Alexander Bengtsson for the inspiring discussions, and the anonymous reviewers for the stimulating questions. This project has been made possible in part by grant number 2020-225832 from the Chan Zuckerberg Initiative DAF, an advised fund of the Silicon Valley Community Foundation. 
\end{acknowledgments}

\section*{Supplemental Material}
Our source codes are available online \cite{kah2022}.

\appendix
\section{Fitting for the remnant field}\label{app:Badd}
The remnant field in the IR chamber is reported to be $B_\text{r}=\SI{42}{\micro\tesla}$ aligned with $+z$ direction.
However, the accurate measurement of such a weak magnetic field in each direction was stated to be a challenge \cite{fris1933ital}.
Therefore, if we allow this remnant field to be a 3-dimensional fitting parameter $\vb{B}_{\text{fit}}=(B_x,B_y,B_z)$, it is possible to improve the model prediction of the FS experimental observation. 
Without hyperfine interaction, the system is reduced to two levels, leading to a closed-form solution similar to the Majorana solution for the quadrupole field \cite{Zener1932, Shevchenko2010, Vitanov1996, ivasn2023}. 
The 2D null point where only the $y$ and $z$ components of the field vanish is located at the coordinates of 
\begin{multline}
   (x'_{\NP},y'_{\NP}, z'_{\NP})=\\ \left(0,\frac{\mu_0 I_\text{w} B_z }{2\pi(B_y^2+B_z^2)}, -z_\text{a} - \frac{\mu_0 I_\text{w} B_y }{2\pi (B_y^2+B_z^2) } \right) \, .
\end{multline}

The total magnetic field in the IR chamber can be approximated \cite{titgk2022} around this null point to the first order as 
\begin{multline} \label{eq:Bquad2}
    \vb{\bm{B}}_{\text{q,fit}} = B_x \,\vux \\
    +\left( \frac{4\pi B_y B_z }{\mu_0 I_\text{w}}(y-y'_{\text{\tiny{NP}}})-\frac{2\pi(B_z^2-B_y^2) }{\mu_0 I_\text{w}}z'_{\text{\tiny{NP}}}\right) \,\vuy \\
    +\left( \frac{2\pi(B_z^2-B_y^2) }{\mu_0 I_\text{w}}(y-y'_{\text{\tiny{NP}}})+\frac{4\pi B_y B_z }{\mu_0 I_\text{w}}z'_{\text{\tiny{NP}}}\right) \, \vuz  \ .
\end{multline}

Using the new approximated field, $\vb{\bm{B}}_{\text{q,fit}}$, the Hamiltonian takes the form
\begin{align}\label{eq:Hfit}
    \nonumber \hat{H}_{\text{fit}}(t) & = - \gamma_\text{e} \frac{\hbar}{2} \, \vb{\bm{B}}_{\text{q,fit}} \cdot \hat{\vb{\bm{\sigma}}} \\
    & = \pmqty{ c_0+c_1 t & a_0-i (b_0+b_1 t) \\ a_0+i (b_0+b_1 t) & -(c_0+c_1 t) } 
\end{align}
with 
\begin{subequations}
    \begin{align}
    a_{0} &= -\gamma_{\mathrm{e}} \frac{\hbar}{2} B_{x} \, , \\
    b_{0} &= \gamma_{\mathrm{e}} \frac{\hbar}{2} \frac{2\pi}{\mu_{0} I_{\mathrm{w}}} \left( 2B_{y} B_{z}y'_{\text{\tiny{NP}}} + (B_{y}^2-B_{z}^2) z'_{\text{\tiny{NP}}}  \right) \, ,\\
    b_{1} &= - \gamma_{\mathrm{e}} \frac{\hbar}{2} \frac{4\pi}{\mu_{0} I_{\mathrm{w}}} B_{y} B_{z} v \, ,\\
    c_{0} &= \gamma_{\mathrm{e}} \frac{\hbar}{2} \left( B_z + \frac{4 \pi}{\mu_0 I_\mathrm{w}} B_y B_z z_\mathrm{a}  \right) \, ,\\
    c_{1} &= \gamma_{\mathrm{e}} \frac{\hbar}{2} \frac{2\pi}{\mu_0 I_\mathrm{w}} (B_y^2-B_z^2) v \, ,
\end{align}
\end{subequations}
where we have replaced $y$ with $vt$ to obtain the time dependence.

To transform into the LZSM formulation, we need to rotate the coordinate system. 
Let us first rotate around $\vux$ by $\theta_x$ using a matrix of $\hat{R}_x=e^{-i\hat{\sigma}_x\theta_x /2}$. This rotation moves all of the time dependence into the diagonal terms. Then, we can rotate around $\vuz$ by $\theta_z$ using a matrix of $\hat{R}_z=e^{-i\hat{\sigma}_z\theta_z /2}$. The final rotated Hamiltonian can be calculated as $\hat{H}_{\text{rot}}(t) = \hat{R}_z\hat{R}_x\hat{H}_{\text{fit}}(t)\hat{R}^{\dagger}_x\hat{R}^{\dagger}_z$. Now, we still need to shift the time coordinate to avoid time-invariant diagonal terms. For this purpose, let us define a new variable $t'=t+t_\text{s}$. 
The rotated and shifted Hamiltonian is given as 
\begin{equation}\label{eq:H1}
\begin{aligned}
\hat{H}_{\text{rot}}(t') &= \hat{R}_z\hat{R}_x\hat{H}_{\text{fit}}(t'-t_{\text{s}})\hat{R}^{\dagger}_x\hat{R}^{\dagger}_z  \\
& = \pmqty{ -\frac{\nu}{2}t' & -\frac{\Delta}{2} \\ -\frac{\Delta}{2} & \frac{\nu}{2}t' } \, ,
\end{aligned}
\end{equation}
with
\begin{subequations}
\begin{align} \label{eq:substitutions1} % {\tiny{\text{Th}}}
\theta_x=&\arctan{\left(\frac{b_1}{c_1}\right)} \, , \\
\theta_z=&\arctan{\left(\frac{b_1c_0-b_0c_1}{a_0\sqrt{b_1^2+c_1^2}}\right)} \, , \\
t_\text{s}=&\frac{b_0b_1+c_0c_1}{b_1^2+c_1^2} \, , \\ 
\nu =& -2\sqrt{b_1^2+c_1^2} \, , \\
\Delta =& -2\sqrt{a_0^2+\frac{(b_1c_0-b_0c_1)^2}{b_1^2+c_1^2}} \, .
\end{align}
\end{subequations}

The solutions to the rotated and shifted system can be given by the parabolic cylinder functions using Zener's solution.
Following the assumptions of adiabatic evolution outside the null point region, the probability of flipping can be found with the Majorana formula $p_{\text{LZSM}}$ in equation \eqref{eq:pLZSM}, where the adiabaticity parameter is 
\begin{multline}
     k=\frac{\Delta^2 }{ \nu \hbar}  = 2\frac{|\gamma_{\mathrm{e}}|}{v} \\ \left( B_y z_{\text{a}} + \left(B_y^2+B_z^2\right)\frac{\pi z_{\text{a}}^{2}}{\mu_{0} I_{\text{w}}} + \frac{B_x^2+B_y^2}{B_y^2+B_z^2} \frac{\mu_{0} I_{\text{w}}}{4 \pi} \right) \, .
\end{multline}
Note that setting $\vb{B}_{\text{fit}}\rightarrow(0,0,B_r=\SI{42}{\micro\tesla})$ yields equation \eqref{eq:k} as expected. 
This probability can be optimized to find the best fit $\vb{B}_{\text{fit}}\approx(16.36,-1.28,53.81)\, \SI{}{\micro\tesla}$. 
The coefficients of determination in this case are $R^2=0.97\, (0.92)$ for the quadrupole (exact) field.

Furthermore, the Rabi equation \eqref{eq:rabi} can be used to obtain a better match. 
Within a reasonable range of fields, the optimized parameters would be $\vb{B}_{\text{fit}}=(0,31.25,66.25)\, \SI{}{\micro\tesla}$ achieving coefficients of determinations $R^2=0.61 \, (0.72)$ for the quadrupole (exact) field.

Because additional remnant fields in the $x$ and $y$ directions were not reported \cite{fris1933,fris1933ital}, achieving a close resemblance through fitting may still fall short of accurately modeling the conditions of the reported experiment.

\section{Modified initial states} \label{app:modrho}

A recently developed theory \cite{wan2022}  matches the experiment well \cite{titgk2022, hetk2022} and yields an anisotropic distribution for the nuclear spin after SG1. 
Anisotropic distributions for the nuclear spin can be obtained if SG1 and the slit do not follow the traditional assumptions made in equation \eqref{eq:maximimallymixed}. 
Inspired by this work, we explore various initial states for the nuclear spin in addition to the maximally mixed state.
We consider an arbitrary mixed initial state
\begin{align}
    \hat\rho_{\text{mixed}}(t_{\text{i}})  &= \sum_{m_I=-I}^{I} d_{m_{\subi} } \ketbra{\psi_{m_{\subi} ,-\sfrac{1}{2} }(t_{\text{i}})}  \ .
\end{align}
The flip probability for such an initial state can be obtained with 
\begin{align}
    p_{\text{mixed}}  &= \sum_{m_I=-I}^{I} d_{m_{\subi} } p_{m_{\subi}}  \ ,
\end{align}
where $p_{m_{\subi} }$ are the flip probabilities if the initial state is a pure eigenstate $ \hat\rho(t_{\text{i}}) = \ketbra{\psi_{m_{\subi} ,-\sfrac{1}{2} }(t_{\text{i}})}$. Due to this relationship, a simple linear regression can be used to fit the experimental observations. Using the standard $a_{\text{exp}}$, the regression is optimized at $(d_{\sfrac{3}{2}},d_{\sfrac{1}{2}},d_{-\sfrac{1}{2}},d_{-\sfrac{3}{2}})\approx(0.285,0,0.643,0.072)$ with an $R^2=0.67 $ for the exact field.

Furthermore, one can consider pure nuclear initial states; however, it is not possible to use linear regression with only $I$ solutions to fit the experimental data. 
One has to solve the von Neumann equation for each initial state considered. 
Various pure states have been tried here; however, the coefficients of determination are not better than that with the mixed state. 

Notwithstanding, modifying initial states for the Frisch--Segr\`e experiment is unconventional in the literature \cite{rab1936}.

\section{Modified HFS coefficients} \label{app:moda}

Up to now, we have used the experimentally measured HFS coefficient value,  $a_{\text{exp}}=\SI{230.8598601(3)}{\mega\hertz}$ \cite{ariiv1977}, which does not accurately predict the experimental observation by Frisch and Segr\`e.
Here, we attempt to improve the match by modifying the hyperfine coefficient.

One way to calculate the HFS coefficient is to use the Fermi contact interaction as follows \cite{fer1930,jac1999, griffiths2018introduction,lev2008}:
\begin{equation}\label{eq:FermiContact}
    2\pi\hbar \ a_{\text{\HFS}} = -\hbar^{2} \frac{2\mu_0}{3} \gamma_{\text{e}} \gamma_{\text{n}}|\psi(0)|^2 \, ,
\end{equation}
where $\psi(\vb{r})$ denotes the spatial wave function of the electron. 
The wave function for the $4\mathrm{s}^1$ electron in $^{39}\mathrm{K}$ does not have an exact solution. 
However, various approximations are available \cite{har1934, ohanian1986spin, wan2022}, yielding the following HFS coefficients: 
% Self average
\begin{subequations}
\begin{align} \label{eq:a_volume} % {\tiny{\text{Th}}}
    a_1 &= -\hbar\frac{\mu_0\gamma_{\text{e}}\gamma_{\text{n}}}{4\pi^2 R^3}\approx \SI{355}{\kilo\hertz} \ , \\
    a_2 &= -\hbar\frac{8\mu_0\gamma_{\text{e}}\gamma_{\text{n}}}{3\pi^4 R^3}\approx \SI{384}{\kilo\hertz} \ , \\
    a_3 &= -\hbar\frac{28.4\mu_0\gamma_{\text{e}}\gamma_{\text{n}}}{6\pi^2 R^3}\approx \SI{6.72}{\mega\hertz} \ , 
\end{align} 
\end{subequations}
where $R=\SI{275}{\pico\meter}$ is the van der Waals radius for $^{39}\mathrm{K}$. Another set of values for $a_{\text{\HFS}}$ are obtained on the basis of an alternative averaging method \cite{wan2022}:
\begin{subequations}
% Torque average
\begin{align} \label{eq:a_torque} % {\tiny{\text{Th}}}
    a_4 &= -\hbar\frac{5\mu_0\gamma_{\text{e}}\gamma_{\text{n}}}{32\pi^2 R^3}\approx \SI{222}{\kilo\hertz} \ , \\
    a_5 &= -\hbar\frac{2\mu_0\gamma_{\text{e}}\gamma_{\text{n}}}{3\pi^4 R^3}\approx \SI{95.9}{\kilo\hertz} \ , \\
    a_6 &= -\hbar\frac{0.138\mu_0\gamma_{\text{e}}\gamma_{\text{n}}}{2\pi^2 R^3}\approx \SI{98.0}{\kilo\hertz} \ .
\end{align} 
\end{subequations}
All of these values along with the experimental value, $a_{\text{exp}}$, have been considered.
Using the maximally mixed initial state, the best match is obtained for $a_{\HFS}=a_{3}$ with $R^2=0.20$ for the exact field. 
Fitting for the initial state using the exact field as in Appendix \ref{app:modrho} yields the best match for $a_{\HFS}=a_{1}$ with $R^2=0.77$. 

\bibliography{ref}

\end{document}